\documentclass[iop,revtex4]{emulateapj}
\usepackage{lineno}

\begin{document}

\renewcommand{\topfraction}{1.0}
\renewcommand{\bottomfraction}{1.0}
\renewcommand{\textfraction}{0.0}

\title{Nearby quintuple systems $\kappa$ Tucanae and $\xi$ Scorpii}

\author{Andrei Tokovinin}
\affil{Cerro Tololo Inter-American Observatory,\footnote{NSF's
    National Optical-Infrared Astronomical Research Laboratory} 
Casilla 603, La Serena, Chile}
\email{atokovinin@ctio.noao.edu}

\begin{abstract}
Architecture and  parameters of  two wide nearby  hierarchical systems
containing five solar-type stars each, $\kappa$ Tuc and $\xi$ Sco, are
studied.  Using  {\it  Gaia}  astrometry and  photometry,  masses  are
determined from  visual orbits and  isochrones, effective temperatures
from  spectra or  colors.  Both  systems are  $\sim$2 Gyr  old. Their
spatial motion corresponds to young  disk but does not match any known
kinematic group.   Internal proper motions  relative to the  center of
mass and radial velocities show that wide $\sim$ 8 kau outer pairs are
bound.  No  correlation  between   orbit  orientations  in  the  inner
subsystems  is observed.  All masses  except one  are confined  to the
narrow range  from 0.8  to 1.5 solar.  Strongly correlated  masses and
wide orbits can be explained  if those systems formed by fragmentation
in  relative isolation and  their components  accreted gas  from common
source, as  expected in a  hierarchical collapse. Young  moving groups
could  be formed  in similar  environments, and  many of  them contain
high-order hierarchies.
\end{abstract} 

\section{Introduction}
\label{sec:intro}

Stellar  hierarchical  systems  with  five or  six  components  occupy
intermediate position between single and  binary stars on one hand and
moving groups and clusters on  the other. Their architecture can throw
light  on the  formation  of these  systems  and therefore  complement
the general  picture of  star formation.  New observations,  in particular
precise  astrometry  from  {\it  Gaia} \citep{Gaia},  allow  study  of
relative  motions  with   unprecedented  accuracy.  Meanwhile,  recent
hydrodynamical        simulations       of        star       formation
\citep{Bate2019,Lee2019,Kuffmeier2019}  provide   details  of  complex
mechanisms involved in the genesis of stellar hierarchies.

High-order hierarchies are rare, but by no means exceptional. Even the
nearest  star,   $\alpha$~Cen,  is  a  triple   system.  According  to
\citet{DK13},  the fraction of  systems with  $N$ components  drops as
$3.7^{-N}$.  The fraction of  triples among  solar-type stars  is 0.13
\citep{FG67},  so  1\% of  all  systems  can  be quintuple,  and  such
hierarchies  are  found  even   in  the  small  sample  within  25\,pc
\citep{R10}. The nearest quintuple systems are GJ~644 (J16555$-$08200)
at 6.4\,pc and $\xi$~UMa (J11182+3122) at 8.3\,pc.  The two quintuples
studied here, $\kappa$~Tuc and $\xi$~Sco, are located at 21 and 28 pc,
respectively, and are composed of solar-type stars.

The  Multiple Star Catalog,  MSC \citep{MSC},  counts 82  entries with
five  or  more components.  Architecture  of  these  systems is  quite
diverse.  Some are  very  young and  contain  pre-main sequence  (PMS)
components.  Many  are members  of  young  moving  groups (YMGs),  for
example   $\alpha$~Gem  (Castor),  $\beta$~Tuc,   $\epsilon$~Cha,  and
$\zeta$~UMa.   \citet{Caballero2010}  discussed  an ultra-wide  (1  pc
separation) multiple  system $\alpha$~Lib and AU~Lib  belonging to the
Castor  group   and  noted  a  relation  between  YMGs  and
hierarchies. He  argued that this system is  gravitationally bound and
not  just a  pair  of group  members.  However, the  borderline
between YMGs and very wide stellar systems remains fuzzy.

High-order  stellar  hierarchies  often  contain  one  or  more  close
spectroscopic subsystems.  For example, there  are three spectroscopic
pairs  in Castor.  However, the  two  quintuples studied  here do  not
contain close subsystems and their wide pairs fit within the canonical
upper  limit of 10  kau. Moreover,  these hierarchies  have an  age of
$\sim$2 Gyr and,  therefore, do not belong to YMGs.  The goal of this
study is  to investigate motions and composition  of these interesting
systems and to propose a scenario of their formation.

Basic information on the architecture of the two selected quintuples and
parameters of their components are given in Section~\ref{sec:obs}, then
each system is discussed in detail in Sections~\ref{sec:KapTuc} and
\ref{sec:XiSco}. Common data and methods are also presented in
Section~\ref{sec:obs}. Formation of these hierarchies is discussed in
Section~\ref{sec:disc}. 

\section{Objects, data, and methods}
\label{sec:obs}

Details on each individual system are given in the two following
Sections. Here, basic information on both systems is assembled,  common
data sources and methods are presented.

\subsection{Structure and parameters of $\kappa$~Tuc and $\xi$ Sco}
\label{sec:quint}

\begin{figure}
\epsscale{1.1}
\plotone{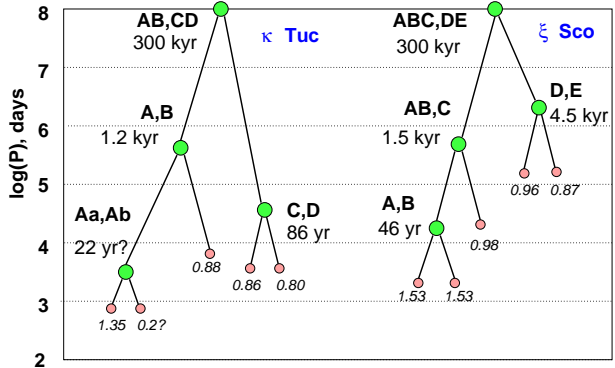}
\caption{Architecture of $\kappa$~Tuc and $\xi$~Sco. The green circles
  represent  subsystems, their  vertical position  corresponds  to the
  logarithm of period  in days, shown on the  vertical axis. The small
  pink circles are individual  stars, with their masses indicated below
  in italics.
\label{fig:mobile} }
\end{figure}

\begin{table*}
\center
\caption{Data on components of $\kappa$~Tuc and $\xi$~Sco}
\label{tab:comp}
\begin{tabular}{l r cc lcc  c  ccc} 
\hline
Comp & HD &  $V$ & $G$ & Sp. & $T_e$ & Mass & $\varpi$ & $\mu^*_\alpha$ &  $\mu_\delta$ & RV \\
     &    & mag & mag & type & K & $M_\odot$  &  mas & mas~yr$^{-1}$ & mas~yr$^{-1}$ & km~s$^{-1}$ \\
\hline
\multicolumn{7}{c}{$\kappa$~Tuc ~~01158$-$6853} & {\bf 47.66} & {\bf 392.7} & {\bf 106.2}  \\
A   & 7788   & 4.88 & 4.79 & F6IV  & 6513 & 1.35 & 47.653 & 409.2 & 107.0 & 8.0 \\ 
B   & \ldots &  7.54 & 7.32 & G5V   & 5145 & 0.88 & 47.528 & 386.3 &  82.4 & 8.2 \\
C   & 7693   & 7.76 & 7.45 & K2V   & 5062 & 0.86 & 47.662 & 360.5 &  95.4 & 5.6 \\
D   & \ldots & 8.26 & 7.94 & K3V   & 4850 & 0.80 & 47.800 & 426.2 & 121.3 & \ldots \\
\multicolumn{7}{c}{$\xi$~Sco ~~16044$-$1122} & {\bf 35.84} & {\bf $-$61.3} & {\bf $-$22.2} \\
A   & 144070 & 4.84 & 4.772 & \ldots  & 6532 & 1.53 & 35.31: & $-$74.3 & $-$32.9 & \ldots \\ 
B   & 144069 & 4.86 & 4.767 & \ldots  & 6532 & 1.53 & 36.24: & $-$41.9 & $-$19.2 & \ldots \\
(A+B) & \ldots &  4.10 & \ldots   & F5IV & \ldots  & 3.06 & 35.776 & $-$58.2 & $-$26.1 & $-$31.18 \\
C   & \ldots  & 7.30 & 7.122 & G1V  & 5705 & 1.00 & 35.822 & $-$75.0 & $-$12.0 & $-$30.35 \\
D   & 144087  & 7.43 & 7.262 & G8V  & 5622 & 0.97 & 35.911 & $-$61.6 & $-$22.2 & $-$31.58 \\
E   & 144088 & 7.99 & 7.784 & K0V  & 5330 & 0.91 & 35.844 & $-$56.4 & $-$20.3 & $-$31.96  \\
\hline
\end{tabular}
\end{table*}

\begin{table*}
\center
\caption{Orbital elements}
\label{tab:orb}
\begin{tabular}{l rrr rrrr rr} 
\hline
System & $P$  & $T$ & $e$ & $a$ & $\Omega$ & $\omega$ & $i$ & $\Sigma M$ & $K_1+K_2$  \\ 
     & yr     & yr  &     & $''$ & $^\circ$ &  $^\circ$ & $^\circ$ & $M_\odot$ & km~s$^{-1}$  \\
\hline
I 27 C,D  & 85.12 & 1916.92 & 0.039 & 1.094 & 141.0 & 135.7 & 31.3 & 1.67 & 4.2  \\
          &$\pm$0.11 & $\pm$1.20 & $\pm$0.002 & $\pm$0.007 & $\pm$0.9 &$\pm$4.6 & $\pm$0.5 & $\pm$0.01 & \ldots \\ 
HJ 3423 A,B & 1200  & 2086.7& 0.40 & 7.03 & 323.1 & 61.3 & 128.5 & 2.22 & 3.1  \\
            & fixed &$\pm$4.1 & fixed & $\pm$0.07 & $\pm$0.5 & $\pm$1.6 & $\pm$1.1 & \ldots & \ldots \\
STF 1998 A,B &  45.864 & 1997.215 & 0.7370 & 0.6666 & 23.86 & 165.61 & 34.37 & 3.06 & 10.1  \\
             & $\pm$0.046 & $\pm$0.012 & $\pm$0.0013 & $\pm$0.0008 & $\pm$0.49 & $\pm$0.56 & $\pm$0.31 & $\pm$0.01 & \ldots  \\
STF 1998 AB,C & 1514 & 2226 & 0.04  & 7.76   & 47.4 & 59.3 & 131.5 &  4.42 & 3.2 \\
\hline
\end{tabular}
\end{table*}

I study here two bright nearby quintuple system composed of solar-type
stars, $\kappa$~Tuc and $\xi$~Sco. These stars have extensive
literature and a good observational history. Nevertheless, only
$\xi$~Sco has been analyzed so far as a quintuple system
\citep{vandeKamp1964,Anosova1991}. Figure~\ref{fig:mobile} shows the
architecture of these systems.  Components (individual stars and centers of mass)
are denoted by one or several letters, systems are designated by
joining their components with comma. For example, A,B refers to the
subsystem containing stars A and B, while AB stands for the center of mass of this
pair in a wider subsystem AB,C.

\begin{figure}
\epsscale{1.1}
\plotone{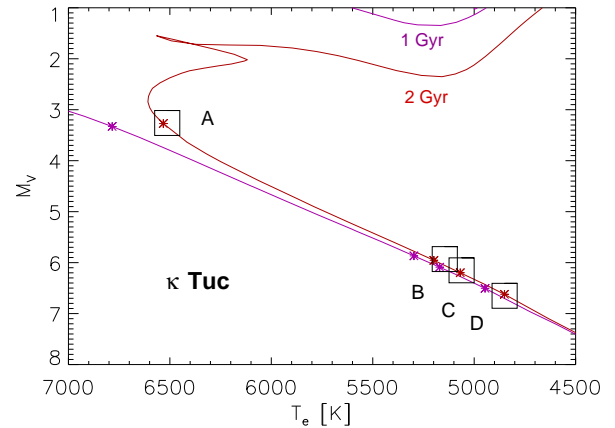}
\plotone{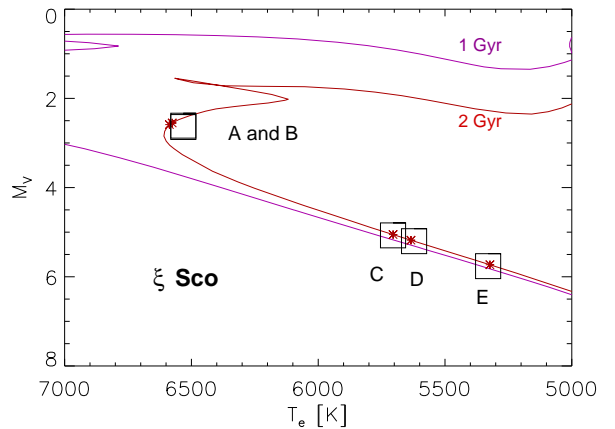}
\caption{ Hertzsprung-Russell diagram  of $\kappa$~Tuc (top) and $\xi$
  Sco (bottom) based on the data of Table~\ref{tab:comp}.  
The magenta and red lines are PARSEC isochrones \citep{PARSEC}
  for solar metallicity and ages 1  and 2  Gyr, respectively.  Asterisks  on the  isochrones
  mark the adopted masses of the components.  
\label{fig:HRD}
}
\end{figure}

Table~\ref{tab:comp} lists data on  the components.   
The adopted parallax of each system and the proper motion (PM)
of the center of mass are given in boldface. Astrometry comes from
{\it Gaia}, radial velocities (RVs), effective temperatures and masses
are discussed below. 
Figure~\ref{fig:HRD} places the components on the
Hertzsprung-Russell diagram (HRD) using data from Table~\ref{tab:comp} and
compares with two solar-metallicity isochrones from \citet{PARSEC}. In
both systems,  massive components are slightly evolved. 

I looked for potentially missed  faint companions using {\it Gaia} and
have  not found  any within  projected distance  of 50  kau  from each
system  down to $M_G \sim 19$  mag, well below the end of the main
  sequence at $M_G  \sim 11$ mag.  This fact,  together with the 
  absence  of  detected  spectroscopic  subsystems,  means  that  all
stellar components are known.

\subsection{Motion of wide pairs}
\label{sec:wide}

Precise  parallaxes and  PMs measured  in the  second {\it  Gaia} data
release   \citep{Gaia}    enable   a   new   look    at   these   wide
hierarchies.  However, knowledge  of component's  masses is  needed to
compute position and motion of the  center of mass of the whole system
or  its  constituents  (subsystems).  Good-quality  visual  orbits  of
subsystems,   re-evaluated  here,  and   precise  parallaxes   of  all
components provide mass measurements with a sub-percent accuracy. They
agree  with stellar  isochrones (Figure~\ref{fig:HRD}),  and therefore
validate  masses   of  other  stars  estimated   from  their  absolute
magnitudes using these isochrones.

Knowing PMs  and masses, I  compute the center  of mass motion  in the
plane of  the sky  as mass-weighted {\it  Gaia} PMs. Long  periods are
estimated from  the third Kepler  law by assuming that  semimajor axis
equals the projected separation $s$.  These periods $P^*$ are valid in
the statistical sense because the  median of the $P^*$ distribution is
close to the  actual period $P$.  The characteristic   speed of
the orbital  motion $\mu^*$ is  computed for a circular  face-on orbit
with a semimajor axis  $s$ \citep{Tok2017}. The actual projected speed
$\Delta \mu$  of a  bound binary cannot  exceed $\sqrt{2}  \mu^*$. The
normalized speed  $\mu ' =  \Delta \mu /\mu^*  $ is distributed  in the
range from  zero to $\sqrt{2}$,  with typical medians around  0.5. The
angle  $\gamma$  between  relative  motion  in a  wide  pair  and  the
radius-vector joining the components also contains some information on
the orbit. The distribution of eccentricities can be inferred from the
joint distribution of $\gamma$ and $\mu'$ \citep{Tok2017}. This cannot
be  done  for  individual  systems,  as  is the  case  here,  but  the
individual values  of $\gamma$ and  $\mu'$ still give some  insight on
the wide orbits.

RVs of the components bring additional
information. They are essential to establish  absence of inner
subsystems and to check that the wide pair is bound, complementing 
 $\mu'$. Unfortunately, the distance between components of a wide
 pair along the line of sight, $\Delta z$, is not known (the {\it Gaia}
 parallaxes are not accurate enough), otherwise a full orbit of a wide
 pair could be computed from its instantaneous position and relative
 velocity. Lacking this information, we may still compute a
 one-dimensional family of possible orbits and thus obtain some
 constraints. This is not done here because the difference of RVs
 between components or subsystems is not known with sufficient
 accuracy.

\subsection{Visual orbits}
\label{sec:wds}

Some  inner subsystems have  published visual  orbits.  I  revise them
here  to compute  accurate masses.  The  mass sum  is proportional  to
$a^3/P^2$ ($a$  is the  semimajor axis and  $P$ is the  period). These
elements are  usually positively  correlated. Therefore, error  of the
mass sum  estimated naively from the  published errors of  $a$ and $P$
would  be too large.  Moreover, weighting  schemes adopted  in fitting
orbits by  different authors lead  to different solutions even  if the
data are  the same. I use  here weights inversely  proportional to the
square of measurement  errors and assign large errors  to the historic
micrometer  measurements,  emphasizing   instead  accurate  data  from
speckle interferometry and space  missions. Orbital elements and their
errors   are   determined  using   the   IDL   code  {\tt   orbit.pro}
\citep{orbit}.  The errors  are checked  by fitting  many  orbits with
randomly perturbed data. This  procedure also delivers realistic error
of the ratio $a^3/P^2$, hence of the mass sum.
 
Positional measurements used in  the orbit fitting were retrieved from
the   Washington   Double  Star   (WDS) Catalog   database  \citep{WDS}   and
complemented  by   recent  speckle  data   \citep{SOAR}  and  relative
positions  measured by  {\it Gaia}.   Elements of  the  updated visual
orbits  and  their  errors  are listed  in  Table~\ref{tab:orb}.   For
completeness,   the   preliminary   orbit   of  STF   1998   AB,C   by
\citet{Zirm2008}  is given in  the last  line. The  penultimate column
gives the  mass sum computed from  $a$, $P$, and  adopted parallax. The
errors of  the mass sum do  not account for errors  of parallax (which
are small for these nearby stars) and potential systematic errors that
are difficult  to evaluate.  For stars with  accurate orbits,  the mass
sums  confirm   mass  estimates  from   the  isochrones.   Conversely,
estimated  masses help  to  constrain uncertain  orbits computed  from
short arcs.  The last column  of Table~\ref{tab:orb} gives the full RV
amplitudes computed from the orbital elements and masses.

\subsection{CHIRON spectroscopy}
\label{sec:chi}

High-resolution ($R\sim  80,000)$ spectra  were taken with  the CHIRON
optical  echelle  spectrometer  \citep{CHIRON}.  The  spectrograph  is
fiber-fed  by  the  1.5  m  telescope  located  at  the  Cerro  Tololo
Interamerican  Observatory (CTIO)  and operated  by Small  \& Moderate
Aperture  Research  Telescope  System  (SMARTS)  Consortium.\footnote{
  \url{http://www.astro.yale.edu/smarts/}}   The   data  analysis   is
described in  \citep{Tok2016}.  A cross-correlation  function (CCF) of
the  reduced spectrum  with a  binary mask  allows us  to  measure the
RV. The  amplitude $A_{\rm CCF}$ and dispersion  $\sigma_{\rm CCF}$ of
the Gaussian  curve approximating the  CCF dip contain  information on
the depth and  width of spectral lines.  The  projected axial rotation
$V \sin  i$ is computed  from $\sigma_{\rm CCF}$ using  calibration in
\citep{Tok2016}.        The      results       are       given      in
Table~\ref{tab:chiron}.  CHIRON spectra  are  discussed below  jointly
with other published spectroscopy.

\begin{table}
\center
\caption{CHIRON observations}
\label{tab:chiron}
\begin{tabular}{l r r r r} 
\hline
Star & JD  & RV  & $A_{\rm CCF}$ & $\sigma_{\rm CCF}$ \\
       & +24\,00000 & km~s$^{-1}$ &    & km~s$^{-1}$ \\
\hline
$\kappa$ Tuc A & 57985.7954  & 8.042   & 0.036 & 31.15 \\
$\kappa$ Tuc B & 57985.7965   & 8.228   & 0.457  &  4.47 \\
$\xi$ Sco AB    & 58922.7639   &  $-$31.179 &  0.107  & 11.00 \\
$\xi$ Sco C     &  58922.7672     & -30.350 &  0.470  &  3.63 \\
$\xi$ Sco D     &  58920.8528  & $-$31.575  &  0.469  &  3.79 \\
$\xi$ Sco E      &  58920.8564 & $-$31.961   & 0.514   &   3.88 \\
\hline
\end{tabular}
\end{table}

\section{$\kappa$ Tucanae}
\label{sec:KapTuc}

$\kappa$~Tuc is composed of  two resolved visual pairs.  The brightest
one  A,B  is $\kappa$~Tuc,  HIP~5896,  HD~7788,  HR~377, GJ~55.3,  and
HJ~3423.   Another pair C,D,  at 318\arcsec  ~($s =  6.7$ kau)  to the
north-west from A,B, is known as HIP~5842, HD~7693, GJ~55.1, and I~27.
This  pair  has  a  reliable  visual  orbit with  a  period  of  85  yr
\citep{Sod1999}.   Both  pairs  have  common WDS  code  J01158$-$6853.
Moreover, the  brightest star A has an  invisible astrometric companion
detected  by  acceleration.   The  average  parallax  of  other  stars
unaffected by  acceleration (47.66 mas, distance 21.0  pc) is adopted.
$\kappa$~Tuc  belongs to  the 25-pc  sample of  \citet{R10},  but they
considered it quadruple because existence of the astrometric subsystem
has not been established at the time.

\subsection{Visual orbits of $\kappa$~Tuc and subsystem Aa,Ab}
\label{sec:visorb}

\begin{figure}[ht]
\plotone{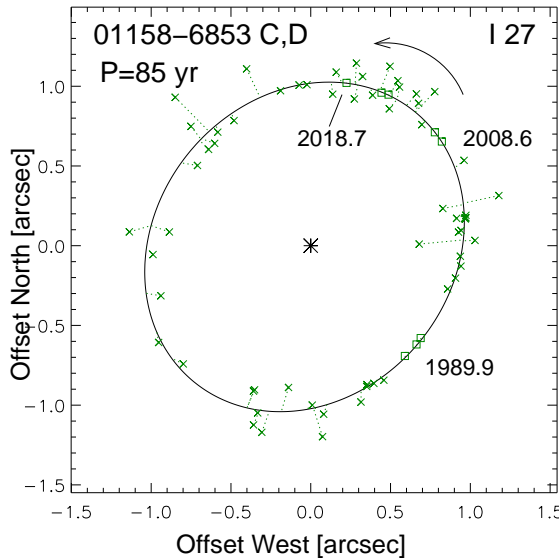} 
\caption{Orbit  of   $\kappa$  Tuc   C,D  (I  27).   Accurate  speckle
  measurements are plotted as  squares, old micrometer measurements as
  crosses.
\label{fig:CD} 
}
\end{figure}

\begin{figure}
\epsscale{1.1}
\plotone{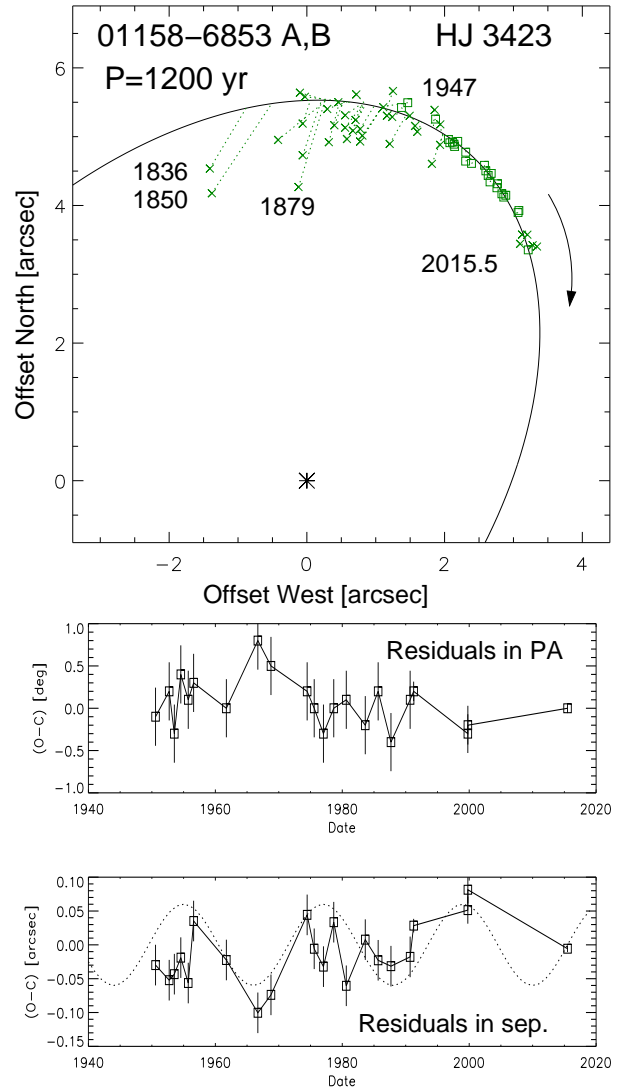}
\caption{Orbit of  $\kappa$~Tuc A,B  (HJ  3423).  Accurate
  photographic  measurements   are  plotted  as   squares,  micrometer
  measurements as  crosses, component A  is at the  coordinate origin.
  The  lower panel  shows residuals  of accurate  measurements  
 in angle and separation. The dotted line is a sine wave with a
  period of 22 yr and an amplitude of 60 mas.
\label{fig:AB1} 
}
\end{figure}

 The orbit of C,D (I~27) by \citet{Sod1999} has a grade 3 in the orbit
 catalog \citep{VB6}. However, the  orbit is accurately defined by the
 new  data obtained  after its  calculation; 1.4  orbital  periods are
 covered since the discovery  of this pair (Figure~\ref{fig:CD}).  The
 corrected  orbital elements  of C,D  and their  errors are  listed in
 Table~\ref{tab:orb}.  The mass sum  is 1.67 $M_\odot$ for parallax of
 47.66 mas.   The relative uncertainly of the  mass sum related to
   the orbit  is $\Delta M/M =  0.006$.  Note  the small eccentricity
 and small inclination  of the orbit to the plane of  the sky, hence a
 nearly constant separation of $\sim$1\arcsec.  The full RV amplitude $K_1
 + K_2$  is only  4.2 km~s$^{-1}$ owing  to the small  inclination and
 long period. Only a small RV  variation of the blended spectrum of CD
 is  expected; blending broadens  the lines  slightly, biasing  the $V
 \sin i$ estimates.

J.~Hershel discovered the pair  $\kappa$~Tuc A,B in 1834.9 at 2\arcsec
~and corrected his first  discrepant measurement to 4\farcs75 in 1836.
The  orbit  with $P=857$  yr  computed  by  \citet{Sca2005} is  poorly
constrained   by  the   short  observed   arc  (Figure~\ref{fig:AB1}).
Adjustment of  the orbit  of A,B is  needed for modeling  its relative
motion  and  searching  for  potential signature  of  the  astrometric
subsystem.   I ignored  a  few most  discrepant micrometer  positions,
assigned errors of 0\farcs5 to the  rest, adopted errors of 30 mas for
photographic positions  available after 1947  and 5\,mas for  the {\it
  Gaia} relative  position. The  weighted residuals confirm  the error
model  ($\chi^2/N  \sim  1$).   To  avoid divergence  and  obtain  the
expected  mass sum  of 2.2  $M_\odot$,  I had  to fix  the period  and
eccentricity.   The new  orbit of  A,B  (Figure~\ref{fig:AB1}) matches
well accurate positions to the detriment of older micrometer data.

According to the adjusted orbit, in  2015.5 star B moved relative to A
with a  velocity of  $(-10.72, -34.68)$ mas~yr$^{-1}$  in RA  and Dec,
respectively.  Subtracting orbital motion from the PM of B, accurately
determined by  {\it Gaia}, I  deduce the average  PM of A  as $(397.0,
117.1)$   mas~yr$^{-1}$.   The   average   PM   of   A   computed   by
\citet{Brandt2018}  from the  difference between  {\it Gaia}  and {\it
  Hipparcos}  positions  is   $(396.6,  117.6)$  mas~yr$^{-1}$.   Good
agreement  between these almost  independent PMs  inspires confidence.
The short-term PMs of A measured by both satellites are different from
the average PM,  and this {\it PM anomaly} $\Delta  \mu$ is a reliable
signature  of the  subsystem. Although  {\it Gaia}  astrometry  of the
bright  star A  is  not very  accurate  (errors $\sim$1  mas), its  PM
anomaly  is highly  significant  and independent  of {\it  Hipparcos}.
According  to \citet{Brandt2018},  $\Delta \mu_{HIP}  =  (14.9, 10.4)$
mas~yr$^{-1}$    and    $\Delta    \mu_{Gaia}   =    (12.6,    -10.6)$
mas~yr$^{-1}$.  During $\Delta  T =  24.25$ yr  elapsed  between these
space missions,  the $\Delta \mu$  vector has turned by  75\degr (from
55\degr  ~to  130\degr)  with  a  roughly  constant  amplitude  of  16
mas~yr$^{-1}$.

Rotation  of $\Delta  \mu$ suggests  a circular  face-on orbit,  and I
adopt this  as a  starting hypothesis to  guess orbital  parameters of
Aa,Ab.  An anti-clockwise  turn by  0.2  fraction of  the full  circle
corresponds to  the orbital  period of  $P = \Delta  T/0.2 =  121$ yr.
However, the  subsystem could have made 0.8  revolutions clock-wise or
1.2 revolutions anti-clockwise  during $\Delta T$ if the  period is 30
or    20    yr,   respectively.     Residuals    in   separation    in
Figure~\ref{fig:AB1}  suggest  a period  $P=22$  yr  that  I adopt  as
plausible.   A sinusoidal  signal with  this period  and  an amplitude
$\alpha = 60$\,mas (dotted line) corresponds to the PM anomaly $\Delta
\mu =  2\pi \alpha/P  = 16$ mas~yr$^{-1}$.   The tentative  sine curve
shows  an  increasing separation  in  2015.5,  hence  A moved  to  the
south-east, in  agreement with {\it  Gaia} $\Delta \mu$.  However, the
residuals in angle do not show  a signal with matching period and
  0\fdg7 amplitude expected for a circular face-on orbit, therefore I
refrain from  fitting an astrometric orbit of  the subsystem. Although
the PM anomaly of star A is highly significant, the period of 22 yr is
only a plausible  guess, and other periods, e.g.  $\sim$120 yr, cannot
be ruled out.

A subsystem  with $P=22$ yr and a  total mass of 1.55  $M_\odot$ has a
semimajor axis of  0\farcs43 and the RV amplitude  $(K_1 + K_2)/ (\sin
i) = 12$ km~s$^{-1}$. The astrometric  amplitude of 60\,mas leads to the mass
ratio $q=0.16$ and $K_1/ (\sin i) = 1.7$ km~s$^{-1}$. A small inclination can
easily  explain the  lack of  detectable  RV variation  caused by  the
subsystem. A companion  Ab of 0.2 $M_\odot$ implied  by this period is
too faint  to be  detectable in the  spectrum or  photometrically, but
could be revealed by high-contrast imaging. If $P \approx 120$ yr, the
companion needed  to produce  the observed PM  anomaly should  be more
massive (0.4 $M_\odot$) and its semimajor axis would be 1\farcs3.

\subsection{Spectroscopy of $\kappa$~Tuc}
\label{sec:rv1}

Table~\ref{tab:rv1}   lists   all   available   RV   measurements   of
$\kappa$~Tuc.  \citet{N04} measured two discordant RVs of A during 4 yr
and  concluded  that  its RV  is  variable  with  an amplitude  of  17
km~s$^{-1}$.  This  result prompted  further  observations  with  the aim  of
determining the spectroscopic  orbit.  \citet{Tok2015} measured RVs of
A,  B, and  CD with  the echelle  spectrograph at  the 2.5  m  Du Pont
telescope. Later, \citet{Tok2015b} used  fiber echelle at the CTIO 1.5
m telescope  and found the  RV of A  to be constant (rms  scatter 0.28
km~s$^{-1}$)  from  10 spectra  taken  during  79  days.  This  excludes  the
short-period  variability.    Another  spectrum  taken   at  the  same
telescope  with CHIRON  in 2017  (Table~\ref{tab:chiron})  extents the
time coverage  to 7 yr  with the same  result, also confirmed  by {\it
  Gaia}  and \citet{Fuhrmann2017}. The  fast axial  rotation of  A ($V
\sin i \approx 60$ km~s$^{-1}$)  prevents accurate measurement of its RV, but
the data  suggest that it changes  slowly, if at  all. The variability
detected   by  \citet{N04}   is  likely   caused  by   one   wrong  RV
measurement. However, the data do not exclude a slow RV variability with
an amplitude of  $\sim$1 km~s$^{-1}$ implied by the  tentative orbit of Aa,Ab
proposed above.

\begin{table}
\center
\caption{Radial velocities of $\kappa$ Tuc}
\label{tab:rv1}
\begin{tabular}{cc  c c l} 
\hline
A & B & CD & Dates  & Reference \\
\hline
1.1?  & \ldots &  6.1 & 1990-s & \citet{N04} \\
7.4  & 8.3 & 5.6   & 2008  & \citet{Tok2015} \\
7.79 & \ldots & \ldots   & 2010  & \citet{Tok2015b} \\
7.48 & 7.91 & \ldots  & 2015.5 & \citet{Gaia} \\
9.4 & 8.25 & \ldots   & 2015.9 & \citet{Fuhrmann2017} \\
8.04 & 8.23 & \ldots  & 2017.6  & CHIRON (this work) \\
\hline
\end{tabular}
\end{table}

Effective temperature of A was  measured at 6513\,K by \citet{Ammler2012}, while
\citet{Fuhrmann2017}   found   $T_e  =   5145   \pm   90$\,K  for   B.
\citet{Ramirez2012} estimated $T_e =  5062 \pm 71$\,K from the blended
spectrum  of CD  and  found  these stars  to  be slightly  metal-rich,
[Fe/H]=0.14.    The  CD   pair  was   monitored  for   exo-planets  by
\citet{Valenti2005}, who give $T_e =  4982$ K and [M/H]=0.05.  I adopt
temperatures of 5062 and 4850\,K for C and D, respectively. The  CHIRON
spectra  of A and  B show   no trace of  the lithium
line and star B rotates slowly.

\subsection{Motion of $\kappa$ Tuc AB,CD}

\begin{figure}
\epsscale{1.1}
\plotone{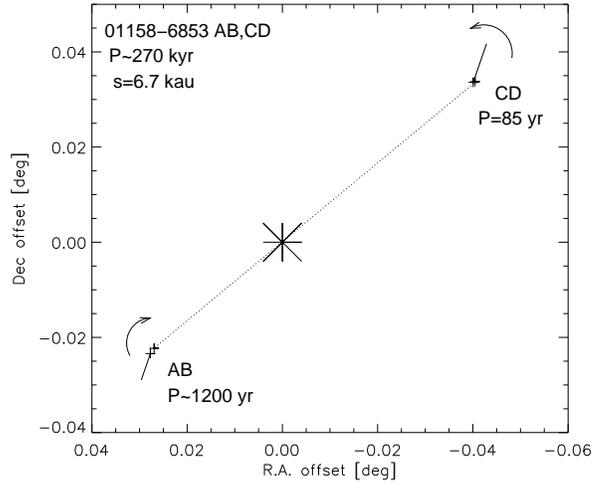}
\caption{Position  of four  resolved  components of  $\kappa$~Tuc on  the
  sky.  Their motion  relative to  the  center of  mass (asterisk)  is
  marked by short lines.
\label{fig:all}  }
\end{figure}

Proper motions  of the  centers of  mass AB and  CD are  determined as
mass-weighted {\it Gaia} PMs, with  the exception of A, where  {\it
  Gaia} PM is  distorted by the subsystem. The PM of  A is computed by
subtracting the computed orbital motion of A,B from the measured PM of
B, the mass of A, 1.55 $M_\odot$, accounts for the astrometric companion.   
Figure~\ref{fig:all} shows  the positions of four stars  in the sky
and the motions  of the two pairs relative to  the common center of
mass. 

CD  moves  relative  to AB  with  the  velocity  $\Delta \mu  =  2.89$
mas~yr$^{-1}$  (0.29  km~s$^{-1}$).   The characteristic  velocity  is
$\mu^* = 7.40$ mas~yr$^{-1}$  (0.74 km~s$^{-1}$), hence the normalized
motion  $\mu' =  0.39$, meaning  that  the system  is bound.  Relative
motion is directed at an angle of $31^\circ$ to the radius-vector, the
separation  between AB  and CD  increases.  Note  that  the subsystems
rotate in  opposite sense (A,B clockwise and  C,D anti-clockwise). The
RV of  CD is biased by  the orbital motion, hence  its difference with
RVs of A and B is not significant.

\subsection{Age and kinematics of $\kappa$ Tuc}
\label{sec:par1}

Figure~\ref{fig:HRD}  places components  of $\kappa$~Tuc  on  the HRD.
The  masses of  C and  D derived  from their  absolute  magnitudes and
isochrone, 0.86 and 0.80 $M_\odot$, match the measured mass sum of CD,
1.67  $M_\odot$.  The  most   massive  star  A  is  slightly  evolved,
suggesting an  age of $\sim$2 Gyr.  Estimates of the age  found in the
literature  are inaccurate  and  discordant, but  none indicates  that
these stars are  young.  The spatial motion $(U,V,W)  = (-34.6, -21.5,
-8.8)$  km~s$^{-1}$ places  this system  among young  disk population.
\citet{Montes2001}  noted similarity  between motions  of $\kappa$~Tuc
and  Hyades, $(U,W,U) =  (-39.7, -17.7,  -2.4)$.  The  multiple system
might  have originated  in  the same  star-formation region,  slightly
enriched in  metals relative  to the Sun,  but it is  definitely older
than the Hyades (otherwise the spectrum of B would contain the lithium
line).


\section{$\xi$ Scorpii}
\label{sec:XiSco}

The system $\xi$ Sco contains five stars in a hierarchical configuration,
recognized as  such by  \citet{vandeKamp1964}.  The main  components A
and B are known  as HR~5978/5977, HD~144070/144069, WDS J16044$-$1122,
STF~1998, and ADS 9909AB.  Another star C is located at 7\arcsec ~from AB.
Further to  the south, at  an angular distance  of 4.7$'$ (8 kau), there  is a
12\arcsec ~pair ADS  9910 (HIP~78738/78739,  HD~144087/144088,  STF 1999)
with common PM,  parallax, and RV (Table~\ref{tab:comp}). 
Although  ADS~9910   has  a  different  WDS   code  16044$-$1127,  its
components belong to the same system and are denoted here as D and E.
Orbital  motion in  the 46-yr  pair A,B  was not  included in  the
{\it Gaia}  astrometric  model,  leading   to  inaccurate  and  oppositely  biased
parallaxes  of A  and B.   However, their  mean,  35.78\,mas, matches
perfectly the  accurate (errors 0.05  mas) parallaxes of  other stars.
The  average  parallax  of  35.84$\pm$0.03  mas  is  adopted  in  the
following (distance 27.9 pc, distance modulus 2.23 mag). 

\subsection{The pair $\xi$ Sco A,B (STF 1998)}
\label{sec:AB2}

\begin{figure}
\epsscale{1.1}
\plotone{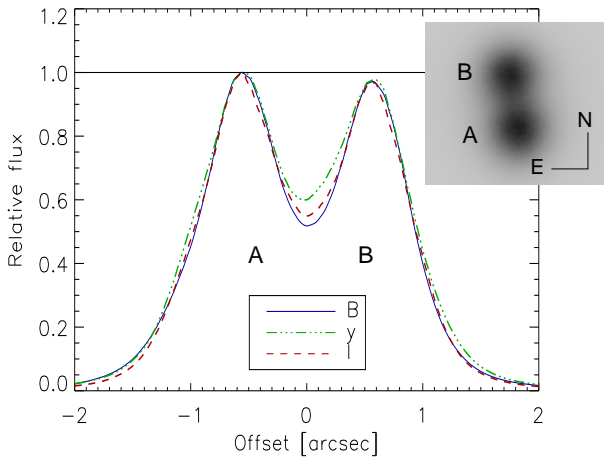}
\caption{Scans of the re-centered average images of $\xi$~Sco AB taken on
  2020-03-13 at SOAR in three filters. The flux is summed perpendicular to the
  binary, normalized by the maximum, and plotted along the separation
  direction. The northern component B is fainter than A by 3$\pm$1\% in all
  filters. The insert shows the average image  of this  pair taken on
  2017-06-06 in the filter  $y$ (532\,nm),   separation
  1\farcs127 and position  angle 7\fdg8. 
\label{fig:AB2} 
}
\end{figure}

The  components A  and B,  presently  at 1\farcs1  separation, are  so
similar that there is some controversy about which of the two stars is
brighter.   The {\it Tycho}  photometry indicates  that B  is brighter
than A  by 0.30 mag in  both $V$ and  $B$ bands, while the {\it  Gaia} $G$
magnitudes of these  stars are practically equal, but  A appears to be
slightly redder  than B.  However, {\it  Gaia} low-resolution slitless
spectroscopy might be seriously  compromised by blending.  It measured
erroneous  effective  temperatures  of  A  and B,  5096  and  5102  K,
corresponding to  spectral type K2.   The combined $V-K$ color  of AB,
1.21 mag,  agrees with  the actual spectral  type F5IV.   However, the
infra-red photometry  of AB in 2MASS  is of poor  quality because the
image is heavily saturated.

The pair A,B is wide enough to be resolved in classical seeing-limited
images, but  no accurate published differential  photometry was found.
It was  observed occasionally by  the speckle camera at  the Southern
Astrophysical Research telescope (SOAR), but its photometry is usually
biased by speckle anisoplanatism  and image truncation in the standard
3\arcsec  ~field.  Figure~\ref{fig:AB2}  shows  an average  re-centered
image   taken  in   2017   in  a   wider   6\arcsec  ~field,   without
truncation. The  experiment was repeated on 2020-03-13  in the filters
$B$,  $y$,  and  $I$.   Scans  through the  images  along  the  binary
convincingly demonstrate that B is  fainter than A by 3$\pm$1\% in all
three filters. Therefore, the colors of A and B are equal.

\begin{figure}[ht]
\plotone{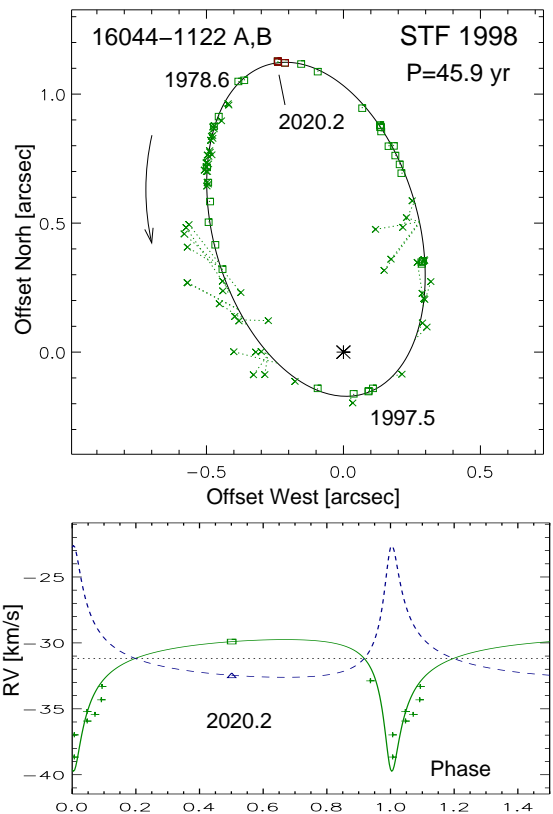} 
\caption{Orbit of STF 1998 A,B. Accurate speckle positions  are plotted as  squares (the
  latest ones in red), selected micrometer data as crosses.  The
  lower plot shows the RV curve with amplitudes  of 5 km~s$^{-1}$. Crosses
  are RVs  from \citet{TS2002}, square  and triangle are  the computed
  RVs in 2020.2.
\label{fig:orbAB} 
}
\end{figure}

\begin{table*}[ht]
\center
\caption{Radial velocities of $\xi$ Sco}
\label{tab:rv2}
\begin{tabular}{cc  cc l} 
\hline
AB & C & D & E  & Reference \\
\hline
\ldots    & \ldots             &  -31.80$\pm$0.30 & -32.30$\pm$0.20 & \citet{N04} \\
-36.3: & -30.90$\pm$0.27 & -31.82$\pm$0.29 & -32.67$\pm$0.36 & \citet{TS2002} \\
\ldots    & \ldots    & -31.79$\pm$0.15 & -32.09$\pm$0.13 & \citet{Gaia} \\
-31.18$\pm$0.1 & -30.3$\pm$0.05  & -31.58$\pm$0.05 & -31.96$\pm$0.05 &  CHIRON (this work) \\
\hline
\end{tabular}
\end{table*}

The orbit  of STF 1998  A,B is well defined  \citep{Doc2009}.  Speckle
interferometry  covers  41.6  yr,   almost  one  full  revolution.  I
re-fitted the  orbit using only speckle data  with appropriate weights
and   the   {\it  Gaia}   relative   position.   Selected   micrometer
measurements  around periastron passages  in 1860  and 1905  are added
with a low weight to  better constrain the period.  The adjusted orbit
is  shown   in  Figure~\ref{fig:orbAB},   its  elements  are   given  in
Table~\ref{tab:orb}.   The  ratio  $a^3/P^2$  is  constrained  with  a
relative  error  of   0.004.   The  parallax  uncertainty  contributes
relative mass-sum error of 0.0025, the contribution of systematic errors is uncertain but likely small. 
The mass sum of AB  is known to better than a percent, 3.06$\pm$0.03
$M_\odot$.  In 2020.2 B was located  to the north of A at a separation
of  1\farcs15, near  the apastron. The  last periastron
passage happened in 1997.2.

\citet{vandeKamp1964} determined  the fractional  mass of B  as 0.479
 based on  photographic astrometry of unresolved pair AB
(motion of the photo-center relative to  reference stars).  As the stars A
and  B  are  very  similar,  the  photocentric  motion  is  small and  its
interpretation  in terms of  mass ratio  depends on the adopted magnitude
difference, which  was quite uncertain at  the time.  The  sign of the
photocentric motion indicates that A is more massive than B.

The orbit predicts motion of B relative to A in 2015.5 with a velocity
of $(+32.33,  +15.42)$ mas~yr$^{-1}$.  The  relative motion measured  by {\it
  Gaia},  $(+32.40, +13.75)$  mas~yr$^{-1}$, agrees  rather  well considering
that these  bright stars  have less accurate  astrometry and  the {\it
  Gaia} linear astrometric solution does not account for the orbit. If
the mean long-term PM of the  pair AB were known, the mass ratio could
be inferred  from the  individual PMs of A and B.  The  average PM of  AB deduced
from the {\it Gaia} PM of C  and the uncertain orbit of AB,C does not
differ significantly from the mean {\it Gaia} PM  of A and B, indicating that the
masses of  A and B are  approximately equal. Indeed,  given the nearly
equal fluxes and colors  of A and B, the mass ratio  in this pair must
be very close to one.

The orbital elements and mass sum correspond to the total RV amplitude
$K_1  + K_2  = 10.1$  km~s$^{-1}$.   Even at  periastron, the  pair is  never
spectrally resolved  because the lines are broadened  by rotation.
  The single-lined  spectroscopic  orbit by  \citet{Chang1929}
with  $K_1 =  3.7$ km~s$^{-1}$  is very  crude.  The  RVs of  AB  measured by
\citet{TS2002} around the 1997.2  periastron correspond to $K_1 = 2.9$
km~s$^{-1}$ and  $\gamma = -33.0$ km~s$^{-1}$. The  fact that the RV  of the blended
spectrum varies so much (60\% of the expected full amplitude) implies
that one of  the components (presumably A) dominates  in the blended spectrum,
possibly  because it  rotates slower  than the  other and  has sharper
lines. The RV of the blended spectrum measured by CHIRON in 2020,
$-31.18$ km~s$^{-1}$,  should be close to the center-of-mass RV. The CCF of AB is slightly asymmetric, but 
an attempt to model it by two Gaussians does not constrain their parameters well enough to be useful.

\subsection{Spectroscopy of $\xi$ Sco}
\label{sec:spec}

These bright  stars has been repeatedly observed by spectroscopists  for various
reasons.  They appear to  be slightly metal-rich and chromospherically
inactive (the  weak X-ray radiation  detected from DE is  explained by
its   proximity to the Sun).   AB   is  normally   not   resolved  in
seeing-limited spectroscopy,  so its spectrum  is a sum of  two nearly
equal stars.  For AB,  \citet{Ramirez2013} give $T_e = 6532$\,K, $\log
g =  4.16$, and [Fe/H]=0.02, while \citet{Casagrande2011}  give $T_e =
6530 \pm 80$K and [Fe/H]=0.11. The catalog of \citet{Hinkel2017} gives
for D and E, respectively, $T_e$ of  5542 and 5245 K, $\log g$ of 4.43
and  4.38, and  [Fe/H] of 0.16  for both.  For the  same stars  D  and E,
\citet{N04} estimated $T_e$ of 5420 and  5164 K and [Fe/H] of 0.06 and
0.09.

Spectra of  the components AB, C, D,  and E were taken  with CHIRON in
2020 March.  In the stars C, D, E the lines are narrow and correspond  to $V \sin i$  of 2.3, 3.0,
and 3.4 km~s$^{-1}$, respectively.  No  lithium line or emissions are seen in
those  spectra.  On the  other  hand,  the CCF  of  AB  is broad  and corresponds
to $V \sin i = 16.9$ km~s$^{-1}$.  In 2020, the RV difference between A and B was only
2.6  km~s$^{-1}$, small  compared to  the rotationally-broadened  lines.  The
lithium  line  6708\AA ~is  present  in the  spectrum  of  AB with  an
equivalent width of 39$\pm$6 m\AA ~and a dispersion of 13.4 km~s$^{-1}$.

The sharp-lined stars  D and E have a  constant RVs over a  long time span
(Table~\ref{tab:rv2}).   The CHIRON spectra  confirm the  RV difference
$\Delta  V_{\rm  ED} =  -0.38$  km~s$^{-1}$,  matching previous  measurements
within  0.1  km~s$^{-1}$.   This  constancy  is  a  strong  argument  against
existence of inner subsystems in D or E.

\subsection{Motion of $\xi$ Sco ABC,DE}
\label{sec:out2}

\begin{figure}
\epsscale{1.1}
\plotone{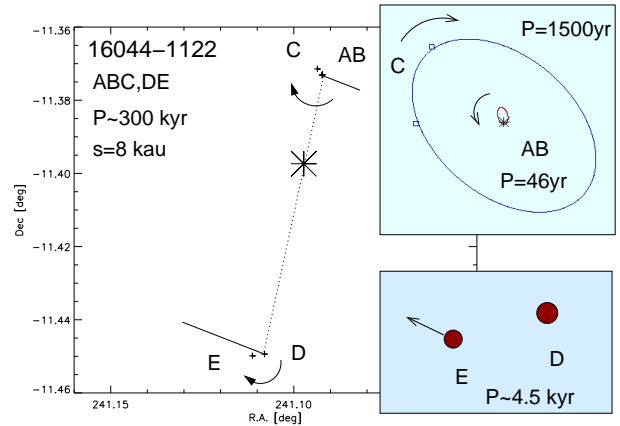}
\caption{ Location  of $\xi$~Sco  on the sky.  The right-hand
  panels  zoom on the  two groups.  The asterisk  marks the  center of
  mass, the lines show  motion of group centers relative to it.
\label{fig:sky2} }
\end{figure}

Figure~\ref{fig:sky2} shows position of the stars on the sky and their
relative  motions determined  as  mass-weighted PMs.   The  PM of  the
center   of  mass   of  the   whole  system   is   $(-61.32,  -22.19)$
mas~yr$^{-1}$.   The   outermost   pair   ABC,DE   rotates   clockwise
(retrograde). Its motion is  almost perpendicular to the radius-vector
joining  ABC  and  DE  and  its  speed  is  3.52  mas~yr$^{-1}$  (0.47
km~s$^{-1}$).   The  projected  separation  $s=8$ kau  corresponds  to
$\mu^* = 6.20$ mas~yr$^{-1}$ or 0.82 km~s$^{-1}$, hence $\mu' = \Delta
\mu/\mu^* = 0.57$. The outer  system is definitely bound and its orbit
likely has a moderate eccentricity because $\gamma \approx 90\degr$.

The intermediate  pair AB,C has  a preliminary (grade 5)  visual orbit
with $P=1514$  yr \citep{Zirm2008}.   The upper-right panel  shows this
poorly constrained orbit determined  from the observed $48^\circ$ arc.
This  orbit corresponds  to  a  plausible mass  sum  and a  retrograde
rotation.  In  contrast,   A,B  has a direct  rotation and
a  large eccentricity.    

The southern pair  D,E has an estimated period of  4.5 kyr. The motion
of  E  relative to  D  is retrograde,  directed  at  an angle  $\gamma
\sim$30\degr ~relative to the radius-vector (E moves away from D). The
speed of  the relative  motion in  the D,E pair  is 5.54  mas~yr$^{-1}$ (0.73
km~s$^{-1}$)  and corresponds  to  $\mu' =  0.33$.  \citet{Brandt2018}  found
marginally significant astrometric accelerations of stars D and E from
comparison  between  {\it  Hipparcos}  and {\it  Gaia}.  This  implies
existence of inner subsystems.   However, {\it Hippacos} astrometry of
double stars  with separations of 10\arcsec--20\arcsec  ~has known problems
caused by its measurement system.  This is the most likely explanation
of  spurious accelerations.  Absence  of subsystems  follows from  the
constant RV difference between D and E.

\subsection{Age and kinematics of $\xi$ Sco}
\label{sec:par2}

Figure~\ref{fig:HRD}        uses        effective        temperatures
(Table~\ref{tab:comp})   to  compare   stellar  parameters   with  the
isochrones. For  D and  E, $T_e$ are  estimated from the  $V-K$ colors
using  the  isochrone.  For  A  and  B, $T_e  =  6532$\,K  is  adopted
\citep{Ramirez2013}, while  for C it corresponds to  the spectral type
G2V. The HRD shows that A and B have evolved off the main sequence,
hence the system is  about 2 Gyr old.  The measured masses  of A and B
match the isochrone almost perfectly. In this region, the isochrone is
vertical, explaining why A  and B have the same color while
differing  slightly in  luminosity.  These stars  are almost  entirely
radiative and  have not fully  depleted lithium in  their atmospheres,
while the less massive stars did.

The  heliocentric spatial  velocity of  the  system is  $(U,V,W) =  (-29.7,  -7.5,   -11.8)$  
km~s$^{-1}$  (U  is  directed   away  from  the Galactic
center). It does  not match any known kinematic  group. The age of
$\sim$2 Gyr implies that the system $\xi$~Sco has been dynamically stable
for a  long time. Hence the  two outer groups  ABC and DE never come
sufficiently close  to each other  to interact dynamically.   This, in
turn, means  that the outer  eccentricity is moderate.  The  motion direction in
the  outer orbit  supports this  view indirectly  (a radial  motion is
expected in an eccentric orbit).

\citet{Anosova1991} studied dynamics  of the triple subsystem ABC and
concluded that it is unstable  with a high probability. However, given
the known orbit of AB,C and the system's age, dynamical instability is
firmly excluded.   These authors  claim that the  system belongs  to a
moving group that  includes ADS~9910 and 12 other  nearby stars listed
in  their Table~8.  I  retrieved modern  data on  those 12  stars from
Simbad  and  computed  their  spatial  motion. The  mean  velocity of
the group (excluding the $\xi$~Sco system) is
$(U,V,W) =  (-30.8, -15.0, -12.8)$  km~s$^{-1}$ and the rms  scatter about
the mean is  $(4.1, 3.6, 3.7)$ km~s$^{-1}$. The mean  $V$ differs from the
velocity of $\xi$~Sco by 7.5 km~s$^{-1}$ (2.1$\sigma$), while the components
$U$ and $W$ are similar. Spatial motions of stars in the Anosova's
table,  as  well  as  of  $\xi$~Sco, correspond  to  the  young  disk
population,  but modern  data do  not provide  evidence of  a putative
kinematic group to which $\xi$~Sco might belong.

\section{Discussion}
\label{sec:disc}

The quintuple  systems $\kappa$~Tuc and $\xi$~Sco  have several common
properties: wide  outer separations $s  \sim 8$ kau, absence  of tight
spectroscopic subsystems, moderate age of $\sim$2 Gyr, and component's
masses distributed  in a  narrow range between  0.8 and  1.5 $M_\odot$
(except  the astrometric companion  $\kappa$ Tuc  Ab). The  {\it Gaia}
catalog indicates  absence of  additional low-mass companions  at wide
separations. In the hindsight, this is not surprising because low-mass
stars with very  wide separations would be torn  apart during lifetime
of these  systems in the  Galactic disk. At  intermediate separations,
there  is  little space  in  the  hierarchy  available for  additional
components  (Figure~\ref{fig:mobile}).    Although  the  dynamical
  stability  criteria allow  such subsystems  to exist  within certain
  ranges of separations, they  could hardly escape detection either by
  {\it Gaia} direct resolution  or by their astrometric signatures, as
  is the  case of $\kappa$~Tuc Aa,Ab.  Inner hierarchical levels
remain free for close subsystems.

Distribution of masses in these systems presents a sharp contrast with
moving  groups and  clusters which  follow the  standard  initial mass
function  where  low-mass stars  dominate.  Dynamical  evolution in  a
cluster could  lead to preferential  binding of more massive  stars in
binaries  while  low-mass  stars  are  ejected.  However,  wide  outer
separations of our hierarchies  strongly speak against origin of these
hierarchies in  clusters. Dissolution of a cluster  could leave behind
wide bound pairs \citep{Kouwenhoven2010},  but it is unlikely
that this process could create high-order hierarchies.

Our hierarchies also appear unusual from the binary-star
perspective. Mass ratio of solar-type binaries is distributed almost
uniformly independently of period \citep{R10,FG67}. Here, all masses
are similar, resembling in this sense low-mass binaries \citep{DK13}. 

Most  likely, these  hierarchies  formed by  core  fragmentation in  a
low-density environment,  in relative  isolation.  As the  gas density
increases during collapse, the Jeans mass decreases, prompting further
fragmentation  that creates  a  hierarchical stellar  system.  At  each
fragmentation,  a   substantial  part  of  the   angular  momentum  of
collapsing  gas is  retained in  the orbital  motion of  the fragments
around common center of mass,  and this storage is most efficient when
the fragment's  masses are  comparable. Therefore, a  two-stage cascade
fragmentation could produce a  wide quadruple system of 2+2 hierarchy,
i.e.  two  close  pairs  on  a  wide orbit  around  each  other,  with
comparable masses.   Such quadruple  systems are indeed  quite common,
e.g. $\epsilon$~Lyr \citep[see further discussion in ][]{Tok2008}. The
architecture of  $\kappa$~Tuc matches  this pattern, except  the inner
subsystem Aa,Ab.

However, in $\kappa$~Tuc the subsystems A,B and C,D rotate in the opposite
sense, their orbits are definitely not coplanar. In $\xi$~Sco, the inner
triple AB,C is misaligned. The angular momenta of the outer and
inner subsystems obviously do not derive from a common source. A quick
study of relative rotation sense in wide 2+2 quadruple systems found
in the MSC confirms that rotation sense of their inner pairs is 
uncorrelated. 

Masses of forming  stars are not determined by  the initial fragments,
but, instead, are set by  continuing accretion of gas for a relatively
long time, compared to the  free-fall time. This happens because small
and dense regions collapse fast,  while collapse at larger scale still
continues.  As  a  result,  in  a  hierarchical  collapse  small-scale
structures    continue   to   accrete    gas   from    larger   scales
\citep{Vasquez2019}.  Components  of a  newly  formed multiple  system
accrete from a common  large-scale gas reservoir, producing stars
with  similar   masses.  Accretion also reduces   the  orbital  separation
\citep{Lee2019,Tok2020}.

Low-density  star  formation  regions  have a  ubiquitous  filamentary
structure, with a typical filament width  of 0.1 pc (20 kau). Gas from
the surrounding cloud falls onto the filament roughly perpendicular to
its axis,  then changes  direction and flows  along the  filament axis
\citep{Vasquez2019,Kuffmeier2019}. Motions  in a dense  clump inside a
filament are  likely directed perpendicular  to its axis,  the angular
momentum  is  hence  roughly   aligned  with  the  filament  axis,  so
fragmentation should  produce a wide pair with  an orbit perpendicular
to  the filament.   However,  the gas  subsequently  accreted by  this
system  comes  along  the  filament,  so  further  fragmentation  into
subsystems and orientation of  their orbits would be uncorrelated with
the outer orbit or mutually.   The accreted gas may contain additional
fragments that fall onto the forming system, get captured by dynamical
friction,  and interact  with other  stars, as  in the  simulations by
\citet{Lee2019}, \citet{Kuffmeier2019}, and \citet{Bate2019}.

The  emerging formation scenario  of $\kappa$~Tuc  and $\xi$~Sco  is a
fragmentation  of  a  relatively   isolated  clump  in  a  low-density
environment, its  growth owing to  prolonged accretion of  gas flowing
from large spatial scales  (along the filament), further fragmentation
into  subsystems,  and  possible  dynamical  interactions  with  other
fragments formed at a  large distance. Comparable masses of components
are  explained by accretion  from  common gas  reservoir. Misaligned
orbits   result    from   random   gas    motions   and/or   dynamical
interactions. However, strong dynamical interactions between stars are
ruled out because otherwise they would have destroyed the weakly bound
outer  pairs. Formation of  wide binaries  by dynamical  ejection from
unstable multiple  systems, proposed  by \citet{RM12}, does  not match
the architecture of these systems.

Stars do not form in isolation.  Other members of the clouds that gave
birth to our hierarchies were  apparently unbound to them (too distant
and with a  too large relative velocity). A  group of mutually unbound
systems is  seen as a YMG  until it disperses.  From this perspective,
the  fact  than  many  YMGs  contain  high-order  hierarchies  appears
natural. It still  remains unclear why our two  systems do not contain
close pairs, unlike other more typical high-order hierarchies.

\acknowledgements

 I thank the Referee, B.~Mason, for pertinent comments.
This work  used the  SIMBAD service operated  by Centre  des Donn\'ees
Stellaires  (Strasbourg, France),  bibliographic  references from  the
Astrophysics Data  System maintained  by SAO/NASA, and  the Washington
Double Star Catalog maintained at USNO.
This work has made use of data from the European Space Agency (ESA) mission
{\it Gaia} (\url{https://www.cosmos.esa.int/gaia}), processed by the {\it Gaia}
Data Processing and Analysis Consortium (DPAC,
\url{https://www.cosmos.esa.int/web/gaia/dpac/consortium}). Funding for the DPAC
has been provided by national institutions, in particular the institutions
participating in the {\it Gaia} Multilateral Agreement.

{\it Facility:}  \facility{CTIO:1.5m}, \facility{SOAR}


\end{document}